\documentclass[ma.tex]{article}

\usepackage[utf8]{inputenc} 
\usepackage[a4paper, total={4.9in, 9in}]{geometry}
\usepackage[T1]{fontenc}    
\usepackage{hyperref}       
\usepackage{url}            
\usepackage{booktabs}       
\usepackage{amsfonts}       
\usepackage{nicefrac}       
\usepackage{microtype}      
\usepackage{graphicx}
\usepackage{amsmath}
\usepackage{amssymb}
\title{\textbf{Student-t Networks for Melody Estimation}}

\author{%
  \textbf{Udhav Gupta, Avi, Bhavesh Jain} \\
  Department of Electrical Engineering\\
  Indian Institute of Technology, Kanpur\\
  \date{May 10, 2021}

}

\begin{document}

\maketitle

\section{Introduction}
The task offered by problem of melody extraction is to obtain the frequency values corresponding to the dominant audio signal in the input monophonic or polyphonic audio data. This problem comprises of many sub-tasks, including, selection of frequency values present as fundamental frequency, figuring out the frequency corresponding to the most dominant voice. \\
The melody extraction problem from audio signals gets complicated when we start dealing with polyphonic audio data. This is because in generalised audio signals, the sounds are highly correlated over both frequency and time domains. This complex overlap of many sounds, makes identification of predominant frequency challenging. \\
In advancement to this, we can deal with filtering background noise in the data, distinguish the components of audio data as voiced and non-voiced, perform instrument recognition. Some mainstream applications are query-by-humming, cover song identification, genre classification, automatic generation of karaoke accompaniment and singer characterization.

\section{Targeted Problem}
Melody estimation or melody extraction refers to the extraction of the primary or fundamental dominant frequency in a melody. This sequence of frequencies obtained represents the pitch of the dominant melodic line from recorded music audio signals. The music signal may be monophonic or polyphonic.\\
\textbf{Monophonic Signal: }Signal is generated from a single source.\\
\textbf{Polyphonic Signal: }Signal is a combination of signals generated by multiple sources simultaneously.
\paragraph{}Melody estimation in a monophonic audio is easier as compared to polyphonic audio. In this paper we have targeted melody estimation for polyphonic audio.
We have compared 2 baseline models based on Signal Processing and patch-based Convolutional Neural Networks in this paper.
We will be implementing Student-Teacher over the 2nd model for learning from less data using a semi-supervised approach with reference to the following paper-
\begin{center}
  \url{https://arxiv.org/pdf/2008.06358.pdf}
\end{center}

\section{Dataset Used}
We have used the dataset MIR-1K for training the model. It contains 1000 audio clips with pitch labeling for window size 40ms and hop size 20ms. Most audio clips are extracted from Karaokes performed mostly by amateurs. Data comprises of male and female audio sounds alike. Unvoiced sounds like friction sounds, inhaling sound, sound of instruments, are also added to the dataset. 
The datasets used for evaluating the models are MIREX-05 and adc2004.

\label{headings}

\section{Methods}
\subsection{Signal Processing}
In this approach we have leveraged the basics of signal processing by computing the auto-correlation of the STFT representation of the input signal. Fundamental frequency is the number of samples in one time period. This time period can be taken to be the distance between the first 2 peaks in the auto-correlation.  

\subsection{Patch based CNN}
We take a novel representation in time-frequency domain i.e CFP. This spectrogram is split into patches of size $25\times 25$. These patches are used as an input to CNN. The CNN comprises of two convolutional layers, followed by three fully connected layers. The convolutional layers have $8, 5\times 5$ filters and $16, 3\times 3$ filters. The number of units in fully connected layers are 128, 64, 2 respectively. Now based on the highest frequency present in a patch, it is classified as either vocal melody(labelled 1) or not (labelled 0). As only small portion of our input data contains vocal melody, data imbalance occurs. Hence, $10\%$ of non-vocal peaks are randomly selected into training data. \\
.The model is trained using the binary cross entropy loss between the CNN output and the ground truth is minimized using batch gradient descent with the Adam optimizer.\\
The CNN outputs the probability of the patch being a Vocal Melody. If the output probability>0.5, then it is considered as vocal melody. Hence, we have the required spectrogram, with vocal melodies set as 1 and others as 0.\\

\begin{figure}[htp]
    \centering
    \includegraphics[width=12cm, height=4cm]{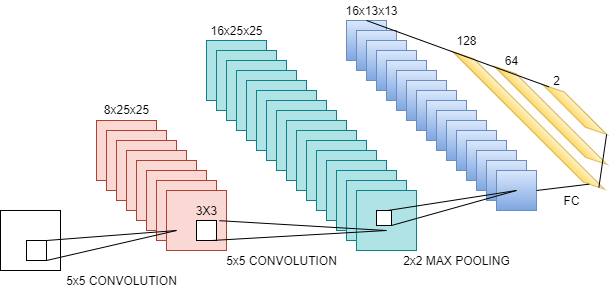}
    \caption{CNN}
\end{figure}

\subsection{Model Architecture}
We propose convolutional recurrent neural network (CRNN) as the baseline architecture. The CRNN architecture consists of 2 ResNet blocks and a bi-directional long short-term memory layer. We first merge the audio waveforms into a mono channel and downsample them to 8 kHz. We then calculate the logarithmic-magnitude spectrogram using short-time Fourier transform with a 1024-point Hann window and an 80-point hop size.The CRNN architecture takes 31 consecutive frames of the spectrogram as input and predicts a pitch label quantized with a resolution of 1/8 semitone. The size of the output layer is 442.

\subsection{Proposed Teacher-Student Models}
Initially we train out Teacher network in a supervised manner i.e with true labels on a dataset $\mathcal{D}$ using the cross entropy loss function.\\
After this we train the Student network in a semi-supervised manner i.e. on a different dataset $\mathcal{U}$, we generate the pseudo labels from our Teacher network and use them as true labels for training the Student network with the following cross-entropy function. 
\[\mathcal{L}_b = \frac{1}{M}\sum_{u=1}^{M} (H(\mathcal{y}_u, p(y|x_u;\Theta_{s})) + H(\mathcal{y}_u, \mathcal{y}_t))\]
Where $\mathcal{y}_u$ is the pseudo labels generated by Teacher Network on $\mathcal{U}$ and $p(y|x_u;\Theta_{s})$ are the predictions made by Student network. And $\mathcal{y}_t$ are true labels. 

\begin{figure}[h]
    \centering
    \begin{minipage}[b]{0.42\textwidth}
        \includegraphics[width = 4 cm,height=5cm]{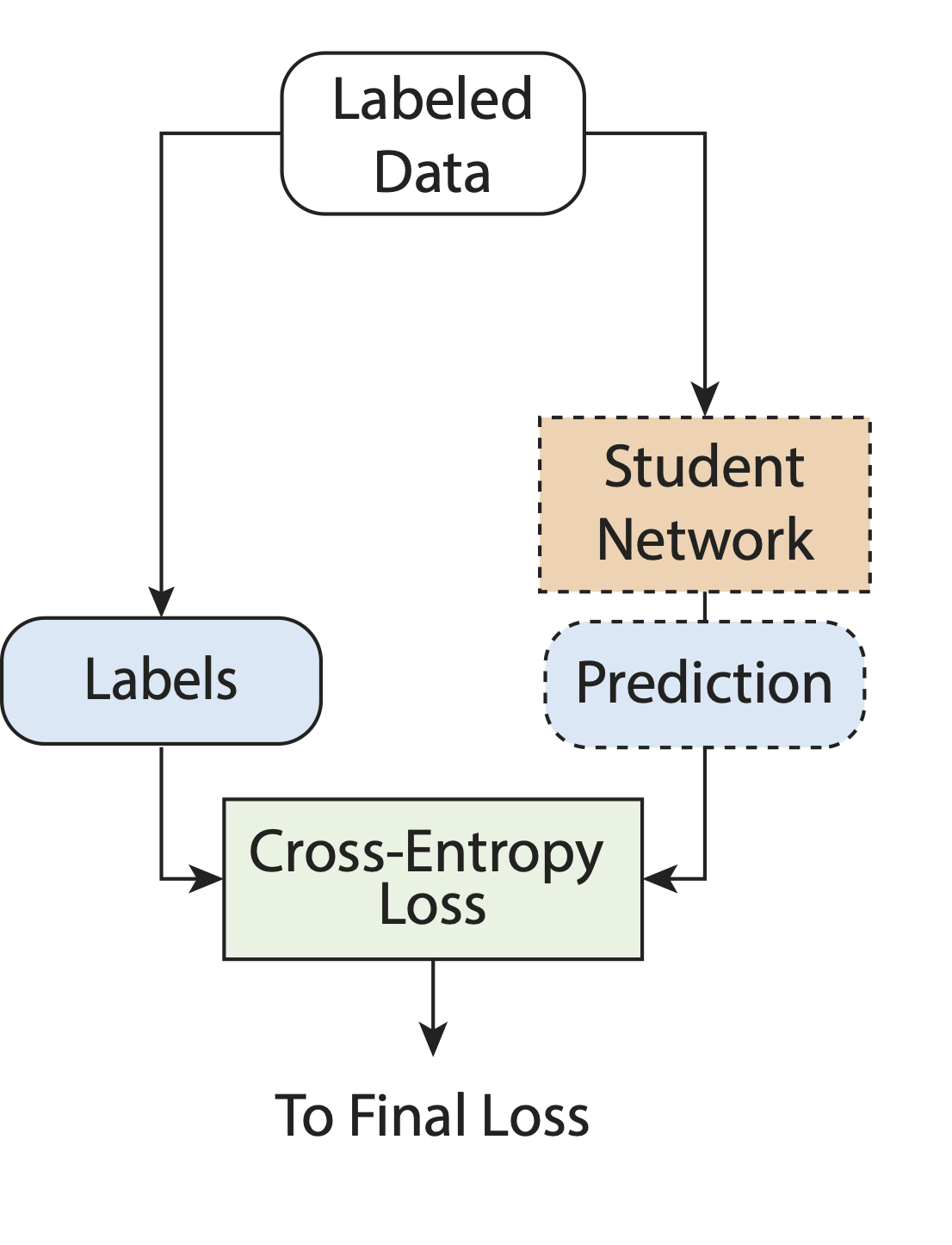}
        \caption{Supervised Loss}
    \end{minipage}%
    \hspace{4.5em}%
    \begin{minipage}[b]{0.32\textwidth}
        \includegraphics[width = 4 cm, height=5cm]{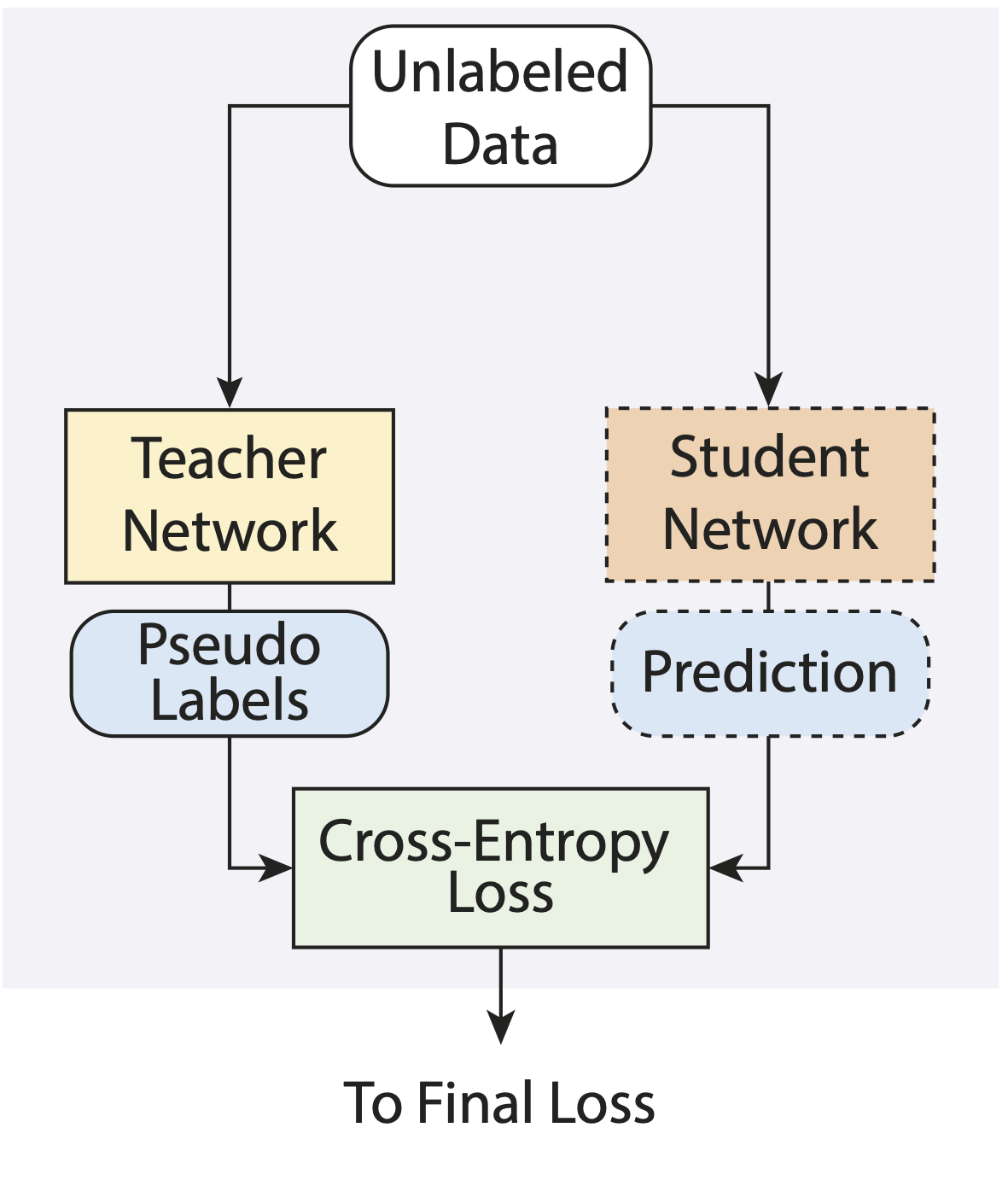}
        \caption{Basic Teacher-Student Loss}
    \end{minipage} 
\end{figure}

\section{Feature Extraction for patch based CNN}
A patch captures events localised in time as well as frequency (or pitch). To effectively localize a pitch event in frequency domain without interference from the harmonics, we use the Combined Frequency and Periodicity representation (CFP). The CNN model proposed in the baseline is then applied on patches selected from this representation. 

\subsection{CFP representation: }
The path-based CNN uses the Combined Frequency and Periodicity (CFP) approach to represent data. This CFP representation is a combination of the Generalized Cepsrtum (GC) and Generalized Cepstrun of Spectogram (GCoS). The GC is obtained by removing the slow-varying portions in the time domain, resulting in sub-harmonics in the lower frequency range. On the contrary, the GCoS majorly represents the harmonics (higher frequency) as the slow varying-portions in the frequency domain are removed.
\textbf{X}:=STFT of input signal amplitude\\
n, k:=Time and frequency indices respectively\\
q:=Quefrency\\
\textbf{W}$_f$, \textbf{W}$_t$:=High Pass Filters\\
\textbf{F}:=N-point Discrete Fourier Transform Matrix\\
$\sigma_i$:=Activation function
\begin{gather*}
    Spectogram: 
    \textbf{Z}_0[k,n]=\sigma_0(\textbf{W}_f\textbf{X})\\
    GC: 
    \textbf{Z}_1[q,n]=\sigma_1(\textbf{W}_t\textbf{F}^{-1}\textbf{Z}_0)\\
    GCoS: 
    \textbf{Z}_2[k,n]=\sigma_2(\textbf{W}_f\textbf{F}\textbf{Z}_1)\\
    CFP:
    \textbf{Y}[p,n]=\tilde{\textbf{Z}}_1[p,n]\tilde{\textbf{Z}}_2[p,n]
\end{gather*}
Before calculating the final CFP representation, we map GC and GCoS to log-frequency domain (similar to pitch). The CFP representation mainly consists of the fundamental frequency as the harmonics and sub-harmonics get suppressed on combining GC and GCoS.  
    
\subsection{Patch Selection}
We assume for every frame in Y, the peak is a vocal melody. We select a patch around each such peak of size 25$\times$25. These patches are the input for the CNN model we have used above.

\begin{figure}[h]
    \centering
    \begin{minipage}[b]{0.42\textwidth}
        \includegraphics[width = 5 cm,height=3cm]{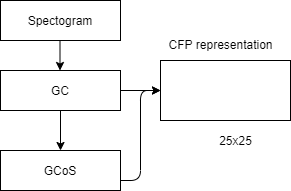}
        \caption{Data Representation}
    \end{minipage}%
    \hspace{4.5em}%
    \begin{minipage}[b]{0.32\textwidth}
        \includegraphics[width = 4 cm, height=3cm]{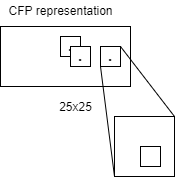}
        \caption{Patch Selection}
    \end{minipage} 
\end{figure}

\section{Results and Conclusions}

\begin{figure}[h]
    \centering
    \begin{minipage}[b]{0.42\textwidth}
        \includegraphics[width = 6 cm,height=5cm]{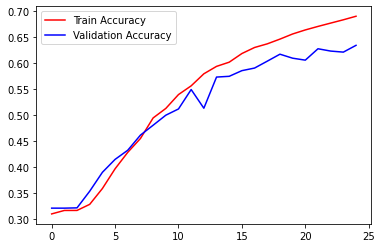}
        \caption{Accuracy for CRNN model}
    \end{minipage}%
    \hspace{4.5em}%
     \begin{minipage}[b]{0.32\textwidth}
        \includegraphics[width = 6 cm,height=5cm]{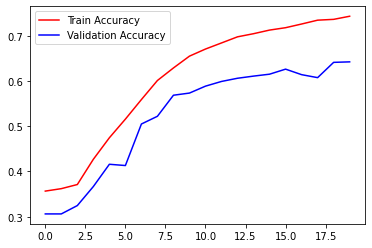}
        \caption{Accuracy for Teacher Student model}
    \end{minipage}%
\end{figure}

\begin{figure}[h]
    \centering
    \begin{minipage}[b]{0.42\textwidth}
        \includegraphics[width = 6 cm,height=5cm]{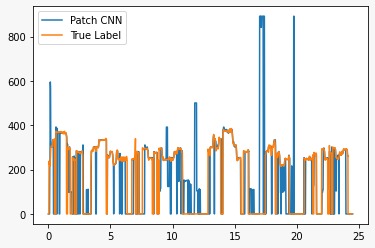}
        \caption{result1}
    \end{minipage}%
    \hspace{4.5em}%
    \begin{minipage}[b]{0.32\textwidth}
        \includegraphics[width = 6 cm, height=5cm]{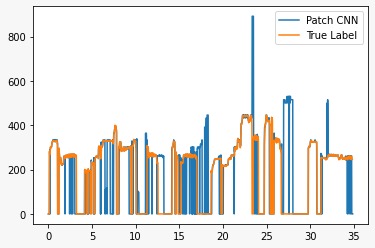}
        \caption{result2}
    \end{minipage} 
   
    \begin{minipage}[b]{0.42\textwidth}
        \includegraphics[width = 6 cm, height=5cm]{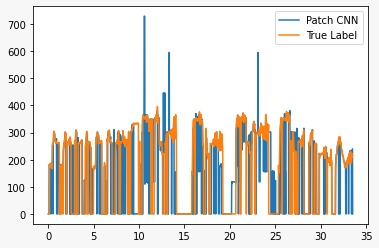}
        \caption{result3}
    \end{minipage}
    \hspace{4.5em}%
    \begin{minipage}[b]{0.32\textwidth}
        \includegraphics[width = 6 cm, height=5cm]{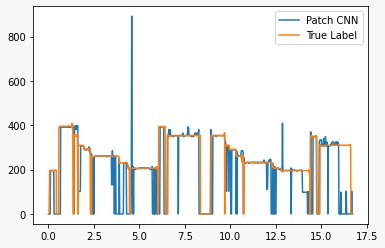}
        \caption{result4}
    \end{minipage}
\end{figure}

\begin{center}
 \begin{tabular}{||c | c | c||} 
 \hline
 Method & ADC2004 & MIREX05  \\ [0.5ex] 
 \hline
 CRNN & 20.29/28.14 &  36.87/45.54 \\ 
 \hline
 Teacher Student & 33.69/41.89 & 65.67/69.92 \\
 \hline
 ResNet NS & 34.44/39.05 & 76.84/79.07 \\
 \hline
 Patch CNN & -/- & 62.86/64.54 \\
 \hline
 SP & 5.01/22.18 & 14.86/34.86\\
 \hline
\end{tabular}
\end{center}

For 2 datasets ADC2004 and MIREX05, from different methods we have evaluated  RPA/RCA values. ResNet RS is the model mentioned in the paper, and hence is used by us for comparison.
\section{Acknowledgement}
We thank Prof. Vipul Arora and all the Teaching Assistants of the course EE698R, 2021 for their constant guidance.

\section*{References}
$[1]$ $http://mirlab.org/dataset/public/$ \\
$[2]$ $https://ieeexplore.ieee.org/document/8462420$ \\
$[3]$ $https://wp.nyu.edu/ismir2016/wp-content/uploads/sites/2294/2016/07/072_Paper.pdf$
$[4]$ $https://ieeexplore.ieee.org/document/8462534$ \\
$[5]$ $https://ieeexplore.ieee.org/document/6739213$ \\
$[6]$ $https://arxiv.org/pdf/2008.06358.pdf$

\end{document}